\documentstyle[twocolumn]{article}
\begin{document}
\twocolumn[\hsize\textwidth\columnwidth\hsize\csname @twocolumnfalse\endcsname
\null\vspace*{2cm}
\centerline{\large Low-Lying Electronic Excitations and Nonlinear Optic Properties of Polymers}
\centerline{\large {\it via} Symmetrized Density Matrix Renormalization Group Method} 
\vspace{0.2cm}
\centerline{ S. Ramasesha$^{1}$, Swapan K. Pati$^{1}$, H. R. Krishnamurthy
 $^{2}$, Z. Shuai$^{3}$ and J. L. Br{\' e}das$^{3}$} 
\centerline{\it ${^1}$Solid State and Structural Chemistry Unit}
\centerline{\it ${^2}$Department of Physics} 
\centerline{\it Indian Institute Of Science, Bangalore 560012 , India.}
\centerline{\it ${^3}$Centre de Recherche en Electronique et
Photonique Mol{\' e}culaires  Service de Chimie des}
\centerline{\it Mat{\' e}riaux Nouveaux  Universit{\' e} de Mons-Hainaut,
 Place du Parc 20, 7000 Mons, Belgium.}

\vspace{0.1cm}
\hbox to \textwidth{\hrulefill}
\leftline{\bf Abstract}

 A symmetrized Density Matrix Renormalization Group procedure together
with the correction vector approach is shown to be highly accurate
for obtaining dynamic linear and third order polarizabilities of
one-dimensional Hubbard and $U-V$ models. The $U-V$ model is seen
to show characteristically different third harmonic generation
response in the CDW and SDW phases. This can be rationalized from 
the excitation spectrum of the systems.

\vspace{0.1cm}


\vspace{0.1cm}
\hbox to \textwidth{\hrulefill}
\vspace{0.3cm}] 

In the electronic structure theory of interacting quasi-one-dimensional
systems, the density matrix renormalization group method (DMRG) has 
established itself as the method of choice. As introduced
by White\cite{white1,white2}, this is a static method for obtaining 
ground and low-lying 
excited states of model systems. We recently extended the method 
to excited states by exploiting symmetries of model systems. 
In this paper, we further extend the method for obtaining accurate 
dynamical linear and nonlinear optic(NLO) coefficients.

The model exact dynamical NLO properties\cite{zyss,keiss,baker} of the 
Pariser-Parr-Pople (PPP) and Hubbard models for finite chains were obtained 
for the first time by Soos and Ramasesha by solving for
correction vectors\cite{soossr}.
The two correction vectors $\phi^{(1)}_{i}( \omega_{1})$
and $\phi^{(2)}_{ij}(\omega_{1}, \omega_{2})$
encountered in the computation of
polarizability and third-order
polarizability are defined by the  following linear equations.
\begin{equation}
({\bf H} -E_{0} + \omega_{1}+
i\epsilon)|\phi^{(1)}_{i}(\omega_{1})={\tilde \mu}_{i}|G>
\end{equation}
\begin{equation}
({\bf H} -E_{0} + \omega_{2}+ i\epsilon)|\phi^{(2)}_{ij}(\omega_{1},
\omega_{2}) ={\tilde \mu}_{i}|\phi^{(1)}_{j}(\omega_{1})>
\end{equation}
\noindent where ${\tilde \mu_{i}}$s are the dipole {\it displacement}
 matrices, $|G>$ and $E_0$ are the ground state eigenfunction and energy,
$\omega$ is the excitation frequency, $\epsilon$ is the damping factor
and ${\bf H}$ is the Hamiltonian matrix.
The correction vector is solved for in the basis of the configuration
functions (Soos and Ramasesha employed a valence bond basis),
which is also the basis
in which the Hamiltonian matrix is set-up for obtaining the ground state.
In terms of these correction vectors, the components
of the polarizabilities,
$\alpha_{ij}$, and third-order polarizabilities, $\gamma_{ijkl}$,
can be written as,
\begin{equation}
\alpha_{ij}(\omega)=<\phi^{(1)}_{i}(\omega)|{\tilde \mu}_{j}|G>+
<\phi^{(1)}_{i}(-\omega)|{\tilde \mu}_{j}|G>,
\end{equation}
\[\gamma_{ijkl}(\omega_{1},\omega_{2},\omega_{3})={\hat P}<\phi^{(1)}_{i}
(\omega_{\sigma})|{\tilde \mu}_{j}|\phi^{(2)}_{kl}(-\omega_{1}-\omega_{2},
-\omega_{1})>\]
\begin{equation}
\end{equation}
\noindent where the operator ${\hat P}$ generates all the permutations:
$( -\omega_{\sigma},i),(\omega_{1}, j), ( \omega_{2}, k)$ and $(\omega_{3},l)$
leading to 24 terms for $\gamma_{ijkl}$ with $\omega_{\sigma} = -\omega_{1}
 -\omega_{2} -\omega_{3}$.
The tumbling averaged
${\bar {\alpha}}$ and ${\bar {\gamma}}$ can be defined as
\begin{equation}
{\bar {\alpha}} = \frac {1}{3} \sum \limits_{i=1}^{3} \alpha_{ii}\;;\quad
{\bar {\gamma}} = \frac {1}{15}\sum \limits_{i,j =1}^{3} ( 2\gamma_{iijj}+ \gamma_{ijji})
\end{equation}
to allow comparison of the calculated NLO response
with experiments on systems containing molecules in random
orientations\cite{zssr}.

\begin{table*}
\begin{flushleft}
Table 1 : Comparison of DMRG and exact $\alpha$ and $\gamma$ values
at $\omega=0.1t$,
$\epsilon=0.001t$ for Hubbard and $U-V$ models of 12 sites. $\alpha$ values
are in units of $10^{-24}$ $esu$ while $\gamma$ values are in units of
$10^{-36}$ $esu$. The $all-trans$ polyene geometry is assumed in the
calculations and bond length is taken as $1 \AA$ for the Uniform chain.
\end{flushleft}
\begin{flushleft}
\begin{tabular*}{17.6cm}{lccccccccc} \hline
NLO \hspace{2cm} & \multicolumn{2}{c} {$U=3t$, $V=0$} \hspace{2cm}& & \multicolumn{2}{c}{$U=3t$,
 $V=1.5t$} \hspace{2cm}& & \multicolumn{2}{c}{$U=3t$, $V=2.0t$} \\ \cline{2-3} \cline{5-6}
\cline{8-9}
Coefficients \hspace{2cm}& Exact & DMRG\hspace{2cm} & & Exact  & DMRG\hspace{2cm} & & Exact & DMRG \\ \hline 
$\alpha_{xx}$\hspace{2cm} & 42.61 & 42.57\hspace{2cm} & & 152.12 & 152.05\hspace{2cm} & & 997.90 & 997.72    \\
$\alpha_{xy}$\hspace{2cm} &  2.95 &  2.91\hspace{2cm} &  & 35.45 &  35.38\hspace{2cm} & & 383.11 & 382.99     \\
$\alpha_{yy}$\hspace{2cm} &  1.52 &   1.44\hspace{2cm} &  & 14.20 & 14.06\hspace{2cm}  & & 154.73 & 154.41   \\
$\bar{\alpha}$\hspace{2cm} & 14.71 &  14.67\hspace{2cm} & & 55.44 & 55.37\hspace{2cm}  &  & 384.21 & 384.04    \\  \hline
$\gamma_{xxxx}$\hspace{2cm} & 26571 & 26566\hspace{2cm} &  & 51380   & 51374\hspace{2cm} & &-32929131 & -32929014  \\ 
$\gamma_{xxyy}$\hspace{2cm} & 113.8 & 109.1\hspace{2cm} & & 2637.7 & 2628.6\hspace{2cm} & &-4915213 & -4915102   \\
$\gamma_{yyyy}$\hspace{2cm} & 1.24  & 1.22\hspace{2cm}  & & 189.9  & 186.0 \hspace{2cm} & &-736663  & -736556     \\
$\bar {\gamma}$ \hspace{2cm}& 5367 & 5356 \hspace{2cm} &  & 11357  & 11345\hspace{2cm}   & &-8704799 & -8704671 \\ \hline
\end{tabular*}
\end{flushleft}
\end{table*}

The above method is quite general and has been employed in the
computations of dynamic NLO coefficients of a wide variety of Hamiltonians
\cite{srzs}. The inhomogeneous linear algebraic equations encountered
in this method  involve large sparse matrices and an iterative small
matrix algorithm, which runs parallel to the Davidson algorithm for
eigenvalue problems, gives rapid convergence for the solution of the
system of equations\cite{sr}.

The DMRG method\cite{white1} as implemented, readily gives the ground state
energy and eigenfunctions and the Hamiltonian matrix. 
The extension of the DMRG method to calculate NLO coefficients by the
correction vector route requires,
in addition, the dipole displacement matrix to be constructed at
each iteration. The matrices of the dipole operators for the
fragments are constructed in the DMRG scheme by renormalizing the matrix
representations of the dipole operator
corresponding to the left and right parts of the system using the density
matrix eigenvector basis in a way which is completely analogous to the
construction of the corresponding Hamiltonian operators
for the fragments. The matrix representation of the dipole operators for
the full system are obtained as direct products of the fragment matrices
analogous to the way by which the full Hamiltonian matrix is constructed.
The dipole displacement matrices are obtained by subtracting the
corresponding components of the ground state dipole moment from the
diagonal elements of the dipole matrices. With the computation of
the dipole displacement matrix, the equations for the correction vector
are completely defined and obtaining the NLO coefficients appears
to be straightforward. However, unless symmetry of the model
Hamiltonian is appropriately exploited, these equations cannot
be solved for frequencies of interest, in most cases.

In the case of model Hamiltonians for polymers, the systems usually possess
total spin symmetry, reflection symmetry about the middle of the chain and
in some cases electron-hole symmetry. These symmetries ensure that a
given correction vector spans only a symmetrized subspace and not the
entire Hilbert space of the Hamiltonian of the given system.
The correction vector lies in the symmetry subspace which is connected
to the ground state by the electric dipole operator. For $\omega$
values corresponding to resonance between the ground state and
eigenstates in that particluar symmetry subspace, $lhs$ of eqns(1) and (2)
become singular for $\epsilon=0$ and present numerical difficulties even
when solving them for reasonable nonzero $\epsilon$ values.
However, if we do not exploit the symmetries of the Hamiltonian,
we encounter singularities in eqns(1) and (2) even  for those
$\omega$ values corresponding to eigenstates of the Hamiltonian found
in other symmetry subspaces which are not connected to the ground
state by the dipole operator. Therefore, numerically it would be impossible
to obtain the correction vectors at these frequencies and thereby the
associated response of the system.

For example, in Hubbard chains at intermediate correlation strengths,
 a triplet excited state lies below the lowest singlet state in the
 {\it ionic} B subspace\cite{srgal}. The states in the {\it ionic} B space
 are connected to the ground state via one-photon transitions. 
The resonances in polarizability are thus expected only at frequencies
corresponding to the energy levels in the {\it ionic} B space,
relative to the ground state. However, we cannot solve for the correction
vector using eqn(1) at an excitation energy corresponding to the 
energy of the lowest triplet state. Thus, the technique of correction
vectors will not be able to give the complete  dispersion of the
polarizabilities upto the first one-photon resonance, unless interferences
due to spurious intruders such as the triplet states are eliminated by
suitably block-diagonalizing the Hamiltonian matrix. This problem of
intruders becomes more severe with increasing system size due to
increasing number of states intruding below the frequency corresponding
to the first "true" resonance.

In the symmetrized DMRG procedure, we have employed Parity (invariance
of Hamiltonian for interchanging second quantized operators of up and
down spins for an even electron system in the $M_s=0$ sector), electron-hole
symmetry and inversion symmetry of the chain. Parity classifies the
spin states as {\it odd}(o) or {\it even}(e) corresponding to odd or
even integer value of total spin.
Electron-hole symmetry bifurcates the states into {\it ionic}(-)
and {\it covalent}(+) spaces and exists only at half-filling. The inversion
symmetry of the chains allows classifing the states as $A_g$ or $B_u$.
 These basic symmetries
are sufficient to exclude spurious resonances in the solution of
eqn(1) and (2) for excitation frequencies in the  range of interest
( usually $\omega \le E_{g}/2$). The implementation of symmetry within
the DMRG procedure has been detailed elsewhere\cite{srskp}

The symmetry adaptation scheme for excited states and NLO properties
has been implemented both
within the infinite chain DMRG and finite system DMRG
algorithms. In the latter, we incorporate the spatial symmetry only when the
left and right parts of the system are identical, {\it i.e.} at the end 
of each finite system iteration. Without iterating over 
the density matrices of the fragments ( i.e. within the infinite system 
algorithm), we find that the energy difference between a chain of length $N$
with $N+1$ and $N-1$ electrons is equal to the Hubbard parameter $U$ to 
an accuracy of $\approx 10^{-3}$. After
three iterations of the 
finite system algorithm, the accuracy improves to $\approx 10^{-5}$, 
for a $U$ value of $4t$.

To illustrate the power of the symmetrized DMRG method, we present results of
DMRG calculations for ground and excited states of uniform Hubbard chains 
at half-filling, for $U/t$ of 4.0 and 6.0, with chain lengths of upto 
50 sites. We have obtained the lowest energy states
in all the eight subspaces of the Schr{\" o}dinger group of the system,
 retaining 100 to 150 eigenvectors of the 
density matrix at each iteration. The extrapolated ground state
energy per site
of the infinite chain for both values of $U/t$
agrees with the exact values of $0.5737331 t$ and $0.4200125 t$ respectively
to 5 decimal places. 
 Model exact gaps vanish for infinite chain.

The energy of the one-photon transition from the ground state
(lying in the $^{e}A^{+}$ subspace) to the lowest energy state in the 
$^{e}B^{-}$ subspace defines the optical gap of the chain. 
The extrapolated DMRG optical gap compares very well with the optical 
gap from the Bethe {\it ansatz} solution for the infinite chain. 
The extrapolated infinite chain values from
a quadratic fit in $1/n$  are $1.278 t$ and $2.895 t$, obtained with 
a cut-off of m=150 for 
$U/t=4.0$ and m=100 for $U/t=6.0$ while the Bethe-{\it ansatz} values are 
$1.2867 t$ and $2.8926 t$ respectively \cite{liebwu}

In fig.1, we present the lowest spin excitation gaps as a function of
inverse chain length for two different values of $U/t$ of 4.0 and 6.0
for uniform Hubbard chains with upto 50 sites. The lowest triplet
state lies in the covalent $A$ subspace but with an odd
parity. The spin gap is defined
as the energy gap between the ground state singlet and the lowest
triplet in the $^oA^+$ subspace. The
Bethe-{\it ansatz} solution yields a vanishing spin gap in the thermodynamic
limit for uniform Hubbard chain at half filling. Second order polynomial
fit in $1/n$ to our DMRG data are consistent with the exact results.

We compare the polarizabilities and third order polarizabilities from the
DMRG calculations with the model exact results for the Hubbard and the
$U-V$ models, in table 1. We find that the DMRG values agree very 
well with the model exact results.

We have computed the dynamic NLO coefficients of the $U-V$ model in the 
SDW ($U >V/2$) as well as CDW ($U \le V/2$) phases with and without
bond alternations. In fig.2 we show the dependence of $ln(\bar \gamma)$
on the number of sites in the chain, $N$ 
(the latter also on a logarithmic scale). For
small $U/t$ and $\delta=0.0$, the dependence exhibits a power-law
behaviour ($\bar \gamma \sim L^{b}, b=3.62$). As we increase $U/t$
or $\delta$, the chain length for which the power law behaviour is seen
is reduced. This shows that the $\pi$-coherence length decreases with
increasing $U/t$ and $\delta$. It is also interesting to note that
in the SDW limit, the dimerized chain ($\delta=0.09$) exhibits a
weaker THG response than uniform chain($\delta=0.0$). A similar behaviour
is also seen for linear polarizabilities\cite{marder}.

In the CDW phase, the optical gap reduces rapidly with chain length.
At moderate chain length ($N \ge 10$) it is observed that the optical
gap is small enough that the first three photon resonance occurs below
$\hbar \omega=0.1 t$ (Optical gap $E_{g}$ becomes $< 0.3 t$)
and this is reflected in a change of sign in all components of
$\gamma$. In the limit of very long chains, the optical gap vanishes
and we should expect $\gamma$ to be negative for any finite but small
excitation frequencies. The behaviour of $\gamma$ in the CDW phase is 
qualitatively different from what is observed in the SDW phase. In the
latter, the optical gap remains finite and the $\gamma$ is expected
to remain positive.

The static $\gamma$ values cannot be computed from correction vector
method as the equations for $\phi_{ij}^{(2)}(\omega_{1}, \omega_{2})$
become singular for $\omega_{2} =0$. However, following Soos and
Ramasesha\cite{soossr}, it is possible to obtain static $\gamma$
values from the dynamic values at two different frequencies by a
Taylor expansion. We have computed $\gamma_{xxxx}(0)$ using this method
 from THG coefficients at two different frequencies of $0.1 t$ and $0.2 t$,
at the crossover region ($U=2V$) between SDW and CDW phases. In fig.3
we present $ln~~ \gamma_{xxxx}(0)$ vs $ln~~ L$ behaviour.
The $\gamma_{xxxx}(0)$
value is positive as the thermodynamic limit of the CDW phase is not
reached even at the maximum chain length (20 sites chain) we have studied.
$\gamma_{xxxx}(0)$ has a power law dependence on chain length L
with an exponent of 3.94. This is very close to the value seen for PPP
models in an earlier exact study of finite chains in static limit\cite{soossr}.

To conclude, the symmetrized DMRG method has been successfully extended
to compute dynamic NLO coefficients. This has been illustrated by
comparison with exact calculation on Hubbard and $U-V$ models. The latter
has been studied in both the CDW and SDW regimes and the qualitatively
different behaviour of $\gamma$ in these two regimes is highlighted.

{\bf Acknowledgement}
It is a pleasure to thank Y. Anusooya for many useful discussions as well
as with help in producing this manuscript. JLB and ZS thank the Belgian
Government SSTC (P\^ole d\'Attraction Interuniversitaire en Chimie
Supramol\'eculaire et Catalyse), FNRS/FRFC and IBM Academic Joint Study.

\begin{center}
Figure Captions
\end{center}
\noindent{\bf Fig.1} \\
Spin Gap(defined in the text) as a function of inverse chain
length 1/N for uniform Hubbard chains with $U/t$ 4.0 and 6.0. Circles
correspond to a DMRG cut-off of $m=100$ and dots to $m=70$. \\
\noindent{\bf Fig.2} \\
Log-log plot of $\bar \gamma$ with site number. The 
model parameters are (i)$U=3t, V=t$(squares), (ii)$U=3t, V=t,
\delta=0.09$(triangles),
(iii)$U=4t, V=t$(diamonds) and (iv)$U=4t, V=t, \delta=0.09$(circles). \\
\noindent{\bf Fig.3} \\
Log-Log plot of $\gamma_{xxxx}(0)$ with chain length
for the parameter $U=3t, V=1.5t, \delta=0.09$.
\end{document}